\documentclass[prb,aps,espf,preprint]{revtex4}
\usepackage{amsmath}
\usepackage{graphicx}
\usepackage{color}

\begin{document}
\title{Magnetic monolayer Li$_{2}$N: Density Functional Theory Calculations }

\author{Gul Rahman}\email{gulrahman@qau.edu.pk}
\author{Altaf Ur Rahman}
\author{Saima Kanwal}
\affiliation{Department of Physics,
Quaid-i-Azam University, Islamabad 45320, Pakistan}
\author{ P. Kratzer}
\affiliation{Faculty of Physics, University of Duisburg-Essen, Lotharstrasse 1, 47057 Duisburg, Germany}

\begin{abstract}
Density functional theory (DFT) calculations are used to investigate the electronic and magnetic structures of a two-dimensional (2D) monolayer Li$_{2}$N. It is shown that bulk Li$_{3}$N is a non-magnetic semiconductor. The non-spinpolarized DFT calculations show that $p$ electrons of N in 2D Li$_{2}$N form a narrow band at the Fermi energy $E_{\rm{F}}$ due to a low coordination number, and the density of states at the Fermi energy ($g(E_{\rm{F}}$)) is increased as compared with bulk Li$_{3}$N. The large $g(E_{\rm{F}}$) shows instability towards magnetism in Stoner's mean field model. The spin-polarized calculations reveal that 2D Li$_{2}$N is magnetic without intrinsic or impurity  defects. The magnetic moment of 1.0\,$\mu_{\rm{B}}$ in 2D Li$_{2}$N is mainly contributed by the $p_{z}$ electrons of N, and the band structure shows half-metallic behavior. {Dynamic instability in planar Li$_{2}$N monolayer is observed, but a buckled Li$_{2}$N monolayer is found to be dynamically stable.} The ferromagnetic (FM) and antiferromagnetic (AFM) coupling between the N atoms is also investigated to access the exchange field strength. {We found that planar (buckled) 2D Li$_{2}$N is a ferromagnetic material with Curie temperature $T_{c}$ of 161 (572) K.}

\end{abstract}
%\pacs{75.30.Et,75.50.Bb,75.30.Kz,77.80.bn,78.20.Bh}

\maketitle

\section{Introduction}
\label{sec:intro}

Currently, huge research efforts are undertaken  
to explore new 2D materials for a wide range of applications.\cite{Sarma,Xu,Mak,Liu,Cai} 
It is well known from semiconductor physics that reduced dimensionality %of a material particularly 
not only affects the electronic structure\cite{elec} and the electronic density of states (DOS) near the Fermi energy, but also the electronic dispersion relations.\cite{book} 
This effect of the dimensionality is particularly strong in 2D materials which, by definition, consist of a sheet only few atomic layers in thickness. 
Due to quantum confinement, the electronic bands in the remaining two in-plane directions can display metallic, insulating, or semiconducting behavior even at variance with the behavior in the 3D bulk compound from which they were derived.\cite{16ref,17ref,18ref,19ref,20ref,21ref,22ref,23ref} In some cases, 2D materials have very unique electronic properties, i.e., Dirac cones in graphene, silicene, etc.\cite{16ref,Kane} {There are also 2D magnetic materials, e.g. transition metal nitrides, where magnetism is due to $d$ electrons. \cite{Cr2017,Cr2017Li} 
Recently, magnetism has been predicted \cite{Cr2016} and confirmed experimentally in a 2D compound of chromium \cite{Cr2017H}. 
In contrast, our present theoretical work predicts the possibility of 2D magnetism even in the absence of $d$ electrons.}

In this paper, we propose 2D materials for applications in the area of spintronics. 
For this purpose, the material must have a large spin polarization at the Fermi energy. 
To date, most of the discovered 2D materials are non-magnetic. 
There have been suggestions to make them magnetic either through intrinsic defects or impurity
atoms. However, due to the typically large formation energies of defects, 
this route appears to be experimentally difficult to pursue.  
Moreover, even if magnetism is experimentally observed, its origin may remain questionable due to the formation
of impurity clusters in the host material that mimic a magnetic signal even in the absence of homogeneous sample magnetization. 
As a way out, we propose to search for alternative 2D materials which show magnetic properties even in the absence of 
intrinsic defects or impurities. 
Therefore, we carried out an extensive computational search using 
density functional theory (DFT) calculations among elemental 2D materials, e.g.  graphene, silicene, germanene, phosphorene, stanene, 
and 2D materials derived from layered bulk compounds,  such as MoS$_{2}$, MoSe$_{2}$, InS, InSe, GaS, and GaSe.\cite{17ref,26ref} 
However, none of them showed any signature of magnetism in their ground states.
The material in focus of this study,  Li$_2$N, is the only material we found 
that possesses spontaneous magnetism without defects. 

With the knowledge gained during this study, 
a guiding principle for the search of magnetic 2D materials can be derived from the general relationship between magnetism and electronic structure 
that can be extended to future work in the field. 
The magnetic exchange interaction, which is crucial for stabilizing a magnetic ground state, is maximum for electrons in atomic orbitals free of radial nodes, such as the atomic $2p$, $3d$, $4f$ orbitals. Keeping the well-known 2D materials in mind, this criterion makes it particularly attractive to search within compounds from the first row of the periodic table, such as carbides and nitrides, that have partially occupied $2p$ orbitals.  
Generally in a 2D material,  
the coordination number of atoms is reduced and/or the electronic wavefunctions are more confined than in the 3D bulk.
Both effects lead to narrower features in the electronic density of states  as compared to bulk. 
For $p$-electron systems, the main focus of the present work, the density of states near the Fermi energy $E_{\rm F}$ will be dominated by $p$ bands. 
In a simple tight-binding picture it can be shown that the bandwidth $W_{p}$ is proportional to the square root of the coordination number. Hence, the bands in a 2D material are generally expected to be narrower than the corresponding bands in 3D bulk. 
If we assume for the sake of simplicity that  $E_{\rm F}$ is located in the band centre, the DOS at the 
Fermi energy ($g(E_{\rm{F}}$)) scales inversely with the band width, $g(E_{\rm{F}})\sim 1/W$.
Thus it is expected that  $g_{p}(E_{\rm{F}})\sim 1/W_{p}$. 
Within the Stoner model of itinerant magnetism, the occurrence of ferromagnetism is expected when  the condition $I\,g(E_{\rm{F}})>1$ is satisfied\cite{28ref}, where $I$ is the exchange integral. Therefore, a large non-magnetic DOS at the Fermi energy and a big $I$ favor ferromagnetism. Hence, our main aim is to search for 2D materials with a narrow band and large $g(E_{\rm{F}}$), resulting in magnetism within the framework of the Stoner model  without  the need to introduce any impurity atoms.
Similar conclusions can be reached also on the basis of the Mott-Hubbard model in case of strongly correlated systems. 
The reduced coordination number in a 2D lattice implies that the electrons have less opportunity to hop from site to site in the lattice. The kinetic energy of the electrons, and hence the electronic band width $W$, will be reduced. 
It is expected that the ratio of $W$ and Coulomb interaction $U$ between the electrons on a given site, $U/W$, moves toward higher Coulomb interactions in reduced dimensionality.  Thus, electron correlation becomes more important and, dependent on the band occupation, the tendencies towards the appearance of magnetism or a Mott transition are enhanced.\cite{mot} 

The DFT results we obtained for a specific material, Li$_2$N, corroborate the above general considerations. 
We stress that Li$_2$N can be logically derived from its bulk parent compound, Li$_3$N, in a similar way as graphene, MoS$_2$, etc. can be derived from their corresponding bulk materials. 
Li$_{3}$N is a superionic conductor 
with several interesting properties and potentials for uses\cite{ref1,ref2,ref3,ref5}. In the past, bulk
Li$_{3}$N was considered to be a host for ferromagnetism induced by transition metals, i.e., Fe,
Co.\cite{35ref,37ref,38ref}  Li$_{3}$N  crystallizes in a hexagonal structure, $\alpha$-Li$_{3}$N, with four ions per unit cell at ambient conditions at equilibrium pressure (Fig.~\ref{structure}). This layered structure consists of Li$_{2}$N layers, widely separated by a pure Li layer which occupies a site between the nitrogen atoms in adjacent layers \cite{ref1}. In the Li$_{2}$N layers each N (0, 0, 0) is at the centre of a regular hexagon formed by the six neighbouring Li ions (1/3, 2/3, 0) and (2/3, 1/3, 0) in units of lattice vectors. 
The unit-cell dimensions are $a = 3.648$\,\AA\,  and $c = 3.875$\,\AA\, with the symmetry point group of $D^{1}6h$ (space group $P6/mmm$)\cite{ref8}. It is clear to see that the hexagonal network of Li and N atoms in Li$_{2}$N layer is very similar to graphene and related 2D materials.

\section{Method of Computation}
%\label{sec:comp}

We performed calculations in the framework of density functional theory (DFT) ~\cite{DFT} using linear combination of atomic orbitals (LCAO) as implemented in the SIESTA code ~\cite{siesta}.
A plane-wave based code QUANTUM ESPRESSO\cite{CM05} was also used.
The local density approximation (LDA)~\cite{lda,CM02} and generalized gradient approximation (GGA)\cite{CM04} were adopted for describing the exchange-correlation interactions.  A cutoff energy of 200 Ry for the real-space grid was adopted in the SIESTA code whereas the wave functions have been expanded into a plane wave basis set with the energy cutoff of 70 Ry.
For the Brillouin zone sampling of the bulk (2D) material, a $20\times 20\times 20$ ($20\times 20\times 1$) $k$-points grid Monckhorst-Pack grid was used in the electronic structure calculations. For 2D-Li$_{2}$N a vacuum slab of 15 \AA\, is used in the direction normal to the  plane of Li$_{2}$N monolayer.
{We considered both the planar and buckled Li$_{2}$N monolayers. The planar Li$_{2}$N monolayer data is presented and we will compare it with the buckled Li$_{2}$N monolayer where necessary.}

\section{Results and Discussion}
\begin{figure}[]
\includegraphics[width=1.0\textwidth]{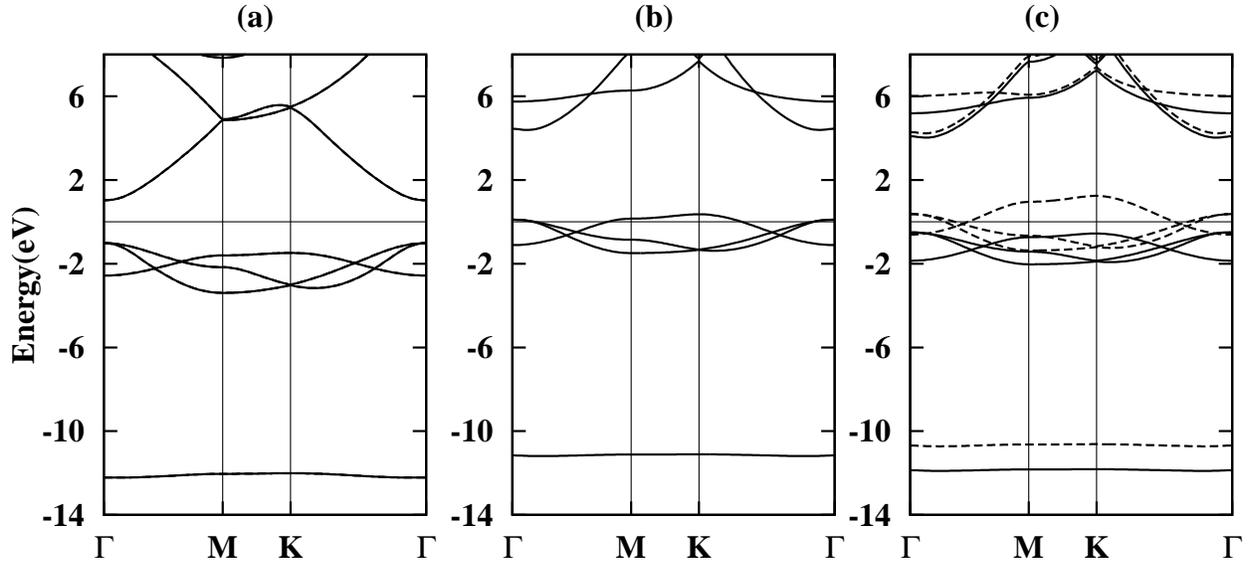}
\caption{(Color online) Calculated band structure of (a) bulk Li$_{3}$N, (b) 2D-Li$_{2}$N nonspin-polarized, and (c) 2D-Li$_{2}$N spin-polarized. The (solid) dashed lines represent the spin-up (spin-down) states.}
\label{bands}
\end{figure}

\begin{figure}[]
\includegraphics[width=0.40\textwidth]{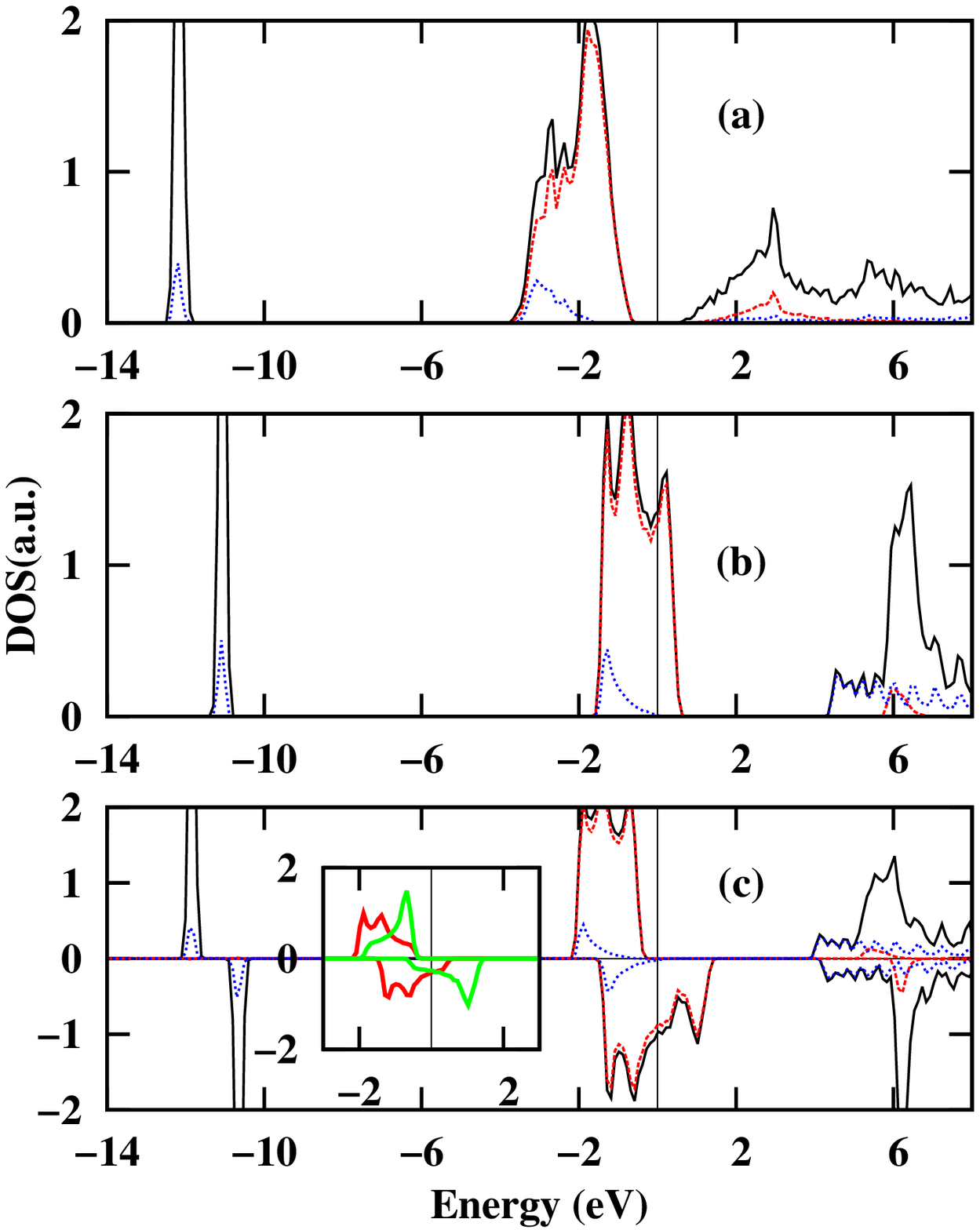} 
\includegraphics[width=0.3\textwidth] {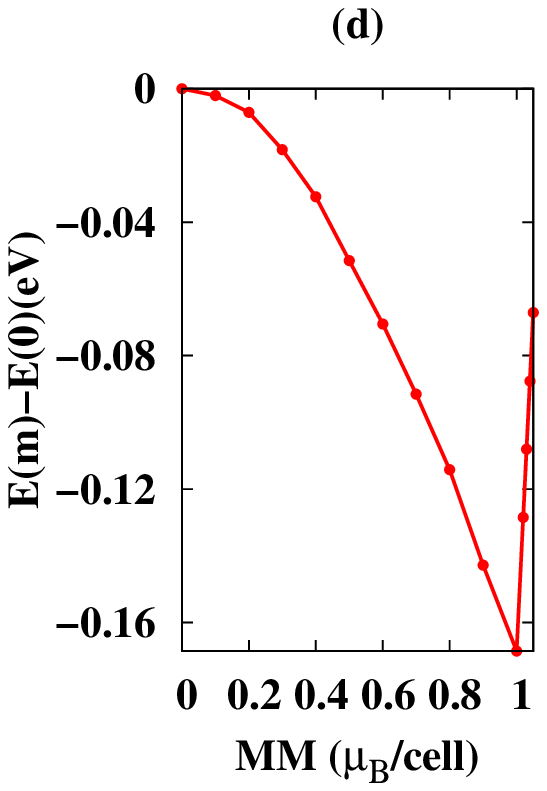}
\includegraphics[width=0.40\textwidth, angle =270] {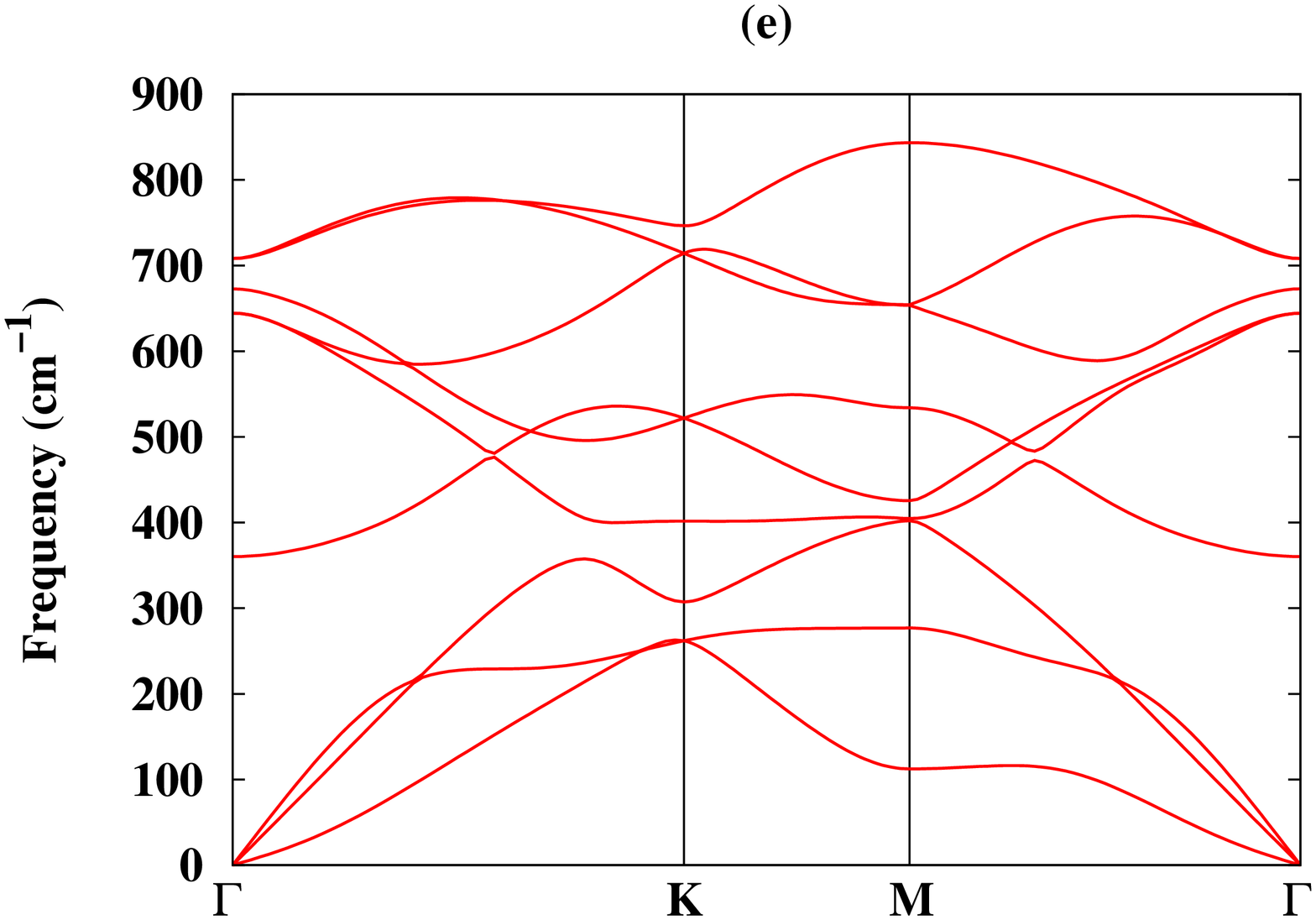}
\caption{(Color online) Calculated total and partial density of states (DOS) of (a) bulk Li$_{3}$N, (b) 2D-Li$_{2}$N nonspin-polarized, and (c) 2D-Li$_{2}$N spin-polarized. The black (solid), red (dashed), and blue (dotted) lines represent the total and PDOS of N-$p$, and Li-$s$, respectively. 
{Positive and negative PDOS values in (c) indicate spin-up and spin-down states.} Inset in (c) represents the $p_{x}$ (solid red) and $p_{z}$ (solid green) orbitals of N. In the inset the vertical line also show the Fermi energy. The calculated total energy (eV) vs magnetic moment MM ($\mu_{\rm{B}}$) is shown in (d). {The phonon dispersion curve of buckled 2D-Li$_{2}$N is shown in (e).} }
\label{PDOS}
\end{figure}

\begin{figure}[]
\includegraphics[width=0.2\textwidth, angle =270]{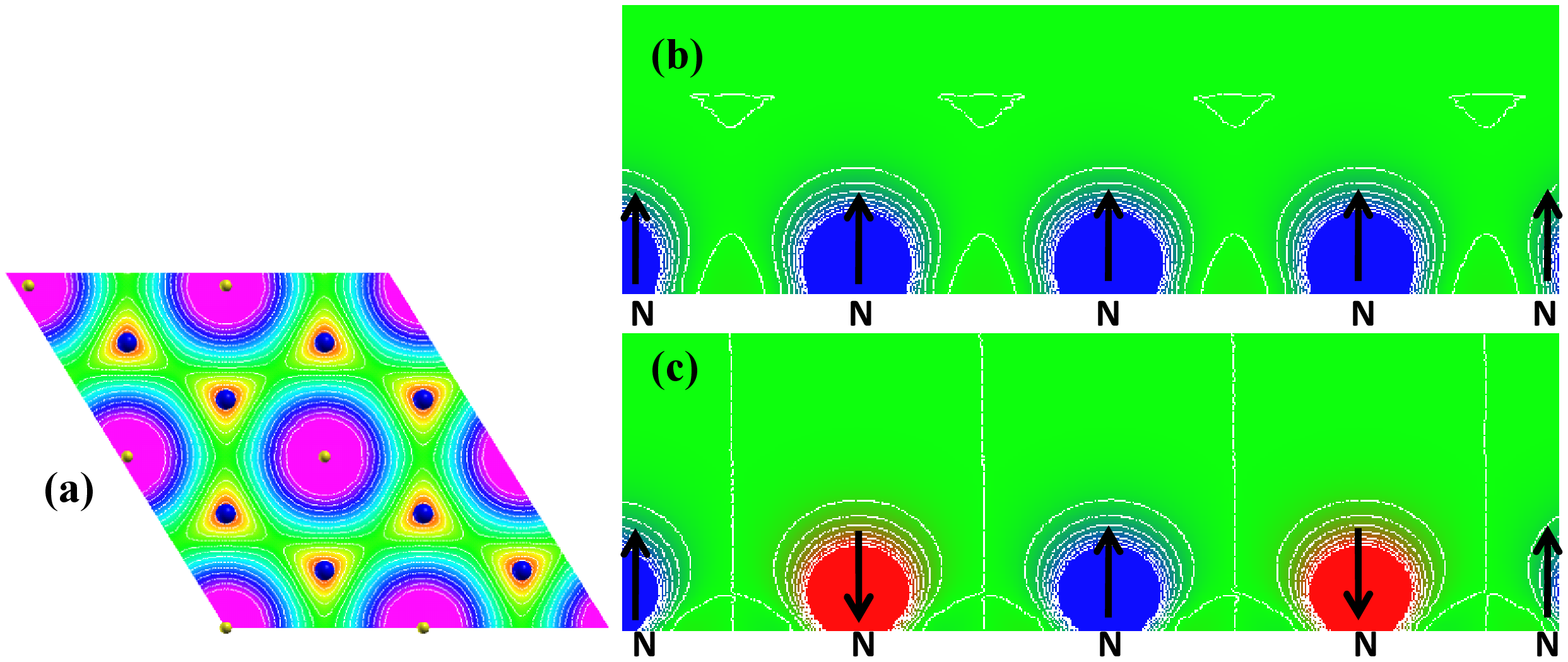}
\caption{(Color online) (a) The
valence electron charge density for Li$_{2}$N in the  (0001) plane. The valence electron charge density in
the (0001) plane. The contours in the
panel is logarithmically spaced. Spin density contours ($\rho_{\uparrow}-\rho_{\downarrow}$) for ferromagnetic (b) and antiferromagnetic (c) coupling of N atoms in 2D Li$_{2}$N. 
Blue (red) colors show positive (negative) spin-polarization. The positive (negative) polarization is represented by spin-up (spin-down) arrows.}
\label{spin}
\end{figure}

%\label{sec:result}
Using the experimental $c/a$ (1.062), our calculated lattice parameter $a$ of bulk Li$_{3}$N is 3.64\AA,\ which is in agreement with the previous work.\cite{ref8,lctheory}
The calculated band structure of bulk Li$_{3}$N is shown in Fig.~\ref{bands}(a). The calculated direct bandgap ($\Gamma$-$\Gamma$) of bulk Li$_{3}$N is $\sim 2.0$ eV, whereas the experimental value is 2.2 eV.\cite{45ref} We also observed that the bandgap of bulk Li$_{3}$N changes with the lattice constant.  A non-interacting isolated flat band around 12eV below the Fermi energy can be seen in Fig.~\ref{bands}(a).  
To probe further the atomic contribution of orbitals, the total and atomic projected density of states of bulk Li$_{3}$N  are also calculated [see Fig.~\ref{PDOS}(a)]. The valance band is mostly dominated by the $p$-orbitals of N and the isolated band at 12 eV is contributed by the N-$2s$ electrons.
The ionic nature of Li$_{3}$N is also visible in the charge density contours, which are shown in
Fig.~\ref{spin}(a). The charge is concentrated around the nitrogen atoms with a radial symmetric
distribution and charge density is nearly spherical around the N atoms. There is also a small
but finite valence density in the neighborhood of the Li atoms which develops a triangular
shape in LiN layer as one moves away from Li core.

To search for spin density, we re-calculated the band structure, charge density of bulk
Li$_{3}$N using the spin-polarized DFT calculations (using both the SIESTA and QE codes),
but no sign of magnetism was observed. However, we observed a magnetic moment when
Li vacancy in bulk Li$_{3}$N was created using large supercells, e.g., $2\times\,2\times2$, 
consistent with the previous work.\cite{39ref} However, the calculated formation energy of Li vacancy in bulk Li$_{3}$N is
large (3.30 eV).  
Motivated by the successful exfoliation of graphite that gives graphene,
we suggest one layer of Li$_{2}$N could be peeled off from the bulk Li$_{3}$N by exfoliation or some other means, e.g. by intercalation or sputtering. We calculated the lattice constant $a$
using both the magnetic (spin-polarized) and non-magnetic (non-spinpolarized) calculations.
The calculated total energy vs lattice constant showed that the spin-polarization slightly
expands the lattice constant by $0.04$\AA\,--similar to other ferromagnets.\cite{40ref,bccfe}
We observed that the magnetic monolayer of Li$_{2}$N has lower energy than the non-magnetic
monolayer of Li$_{2}$N. The magnetic behavior is also confirmed using the GGA and QE code
{The buckling parameter, which is the height $z$ from the N atoms, is found to be $\sim 0.57$\AA. The calculated binding energy also shows that the buckled Li$_{2}$N is more stable than the planar Li$_{2}$N by 0.27 eV}

The lift-off of Li$_{2}$N from the bulk Li$_{3}$N will not only change the chemistry but also the coordination number of nitrogen. Note that N is surrounded by eight Li ions in bulk Li$_{3}$N, but it is surrounded by six Li ions in monolayer of Li$_{2}$N. Due to ionic nature of bulk Li$_{3}$N, no uncompensated spins are present for magnetism. However, when Li$_{2}$N is  lifted-off, the oxidation state of N is changed from  N$^{3-}$ to N$^{2-}$ based on the charge neutrality condition, and following the Hund's rules an uncompensated spin is expected that will lead to local magnetic moments in Li$_{2}$N. We calculated the band structure of Li$_{2}$N in the non-spinpolarized (non-magnetic NM) case [see Fig.~\ref{bands}(b)], and  it is interesting to see that the band becomes narrower (as expected from the discussion in the introduction). 
Some of the N-$p$ states are partially occupied. We also calculated the DOS of Li$_{2}$N in the NM state as shown in Fig.~\ref{PDOS}(b). One can clearly see large density of states at the Fermi energy ($g(E_{\rm F}) \sim 1.50$ states/eV) consistent with our speculation that reduced coordination number will increase the electronic states at the Fermi energy and will fulfil the Stoner condition for ferromagnetism. Our detailed analysis shows that these are the N-$p$ orbitals that increase the DOS at the Fermi energy, the 
N-$s$ orbitals are completely occupied. {Note that the formation of a narrow band at the Fermi energy in Li$_{2}$N is different from other 2D materials~\cite{Sarma,Xu,Mak,Liu,Cai,Topsakal} mainly due to the $p_z$-orbitals of N 
forming $\pi$ bonds.}

We further considered the spin-polarized (magnetic) band structure of Li$_{2}$N as shown in Fig.~\ref{bands}(c). Large spin-polarization deep in the valance band and near the Fermi energy can be seen. An exchange splitting of about 1.2 eV deep in the valance band is visible. The
spin-down (minority) band is wider than the spin-up (majority) band.
The bands are fully spin-polarized, where one spin (spin-up) behaves
as an insulating whereas the other spin (spin-down) is showing metallic behavior, and such
a spin-polarized band structure is important for spintronic devices. {The half-metallic band structure of Li$_{2}$N is similar to other $d^{0}$-electron systems, e.g functionalized silicenes.\cite{zhang2012,zhang2012NRL}} To gain further insight
into the origin of magnetism and spin polarization, the total and atom-projected density
of states of Li$_{2}$N are shown in Fig.~\ref{PDOS}(c). One can easily judge that the spin-polarization is
mainly caused by the N-$p$ orbitals which are completely occupied in the majority spin states
and partially occupied in the minority spin states giving a metallic behavior, and hence
half-metallic band structure. The exchange splitting $I$ of the N-$p$ orbitals near the Fermi
energy is about $1.70$ eV, and hence $Ig(E_{\rm{F}})>1$ is satisfied. The band width of the spin-up
(spin-down) band is about 2.0 (3.0) eV. The N-$p$ and Li-$s$ orbitals hybridization near the
Fermi energy can be seen. As the N-$p$ orbitals consist of $p_{x}, p_{y}, p_{z}$, it is 
interesting to see how these orbitals are polarized and contribute to magnetism in Li$_{2}$N. In the inset of
Fig.~\ref{PDOS}(c), the $p_{x}/p_{y}$ orbitals are mostly occupied in both the spin-states, however, the N-$p_{z}$ orbitals are completely occupied in the spin-up states and nearly empty in the spin-down
states. Therefore, the local magnetic moment of N in Li$_{2}$N is mainly contributed by the $p_{z}$
orbitals, which is also confirmed by our detailed analysis of Mulliken orbital populations. Note that the total magnetic moment of 2D-Li$_{2}$N is 1.0\,$\mu_{\rm{B}}$ per unit cell.
{To further confirm whether the magnetism of 2D-Li$_{2}$N is affected by quantum many-body effects, we carried out additional test calculations using $G_{0}W_{0}$ approximation as implemented in the Yambo code.\cite{yambo} The band structure of Li$_{2}$N calculated with $G_{0}W_{0}$ also shows half-metallic ferromagnetism. The buckled Li$_{2}$N retains its ferromagnetic ground state structure and has 1.0$\mu_{\rm{B}}$ magnetic moment per unit cell.}

\begin{figure}[]
\includegraphics[width=0.25\textwidth, angle =270]{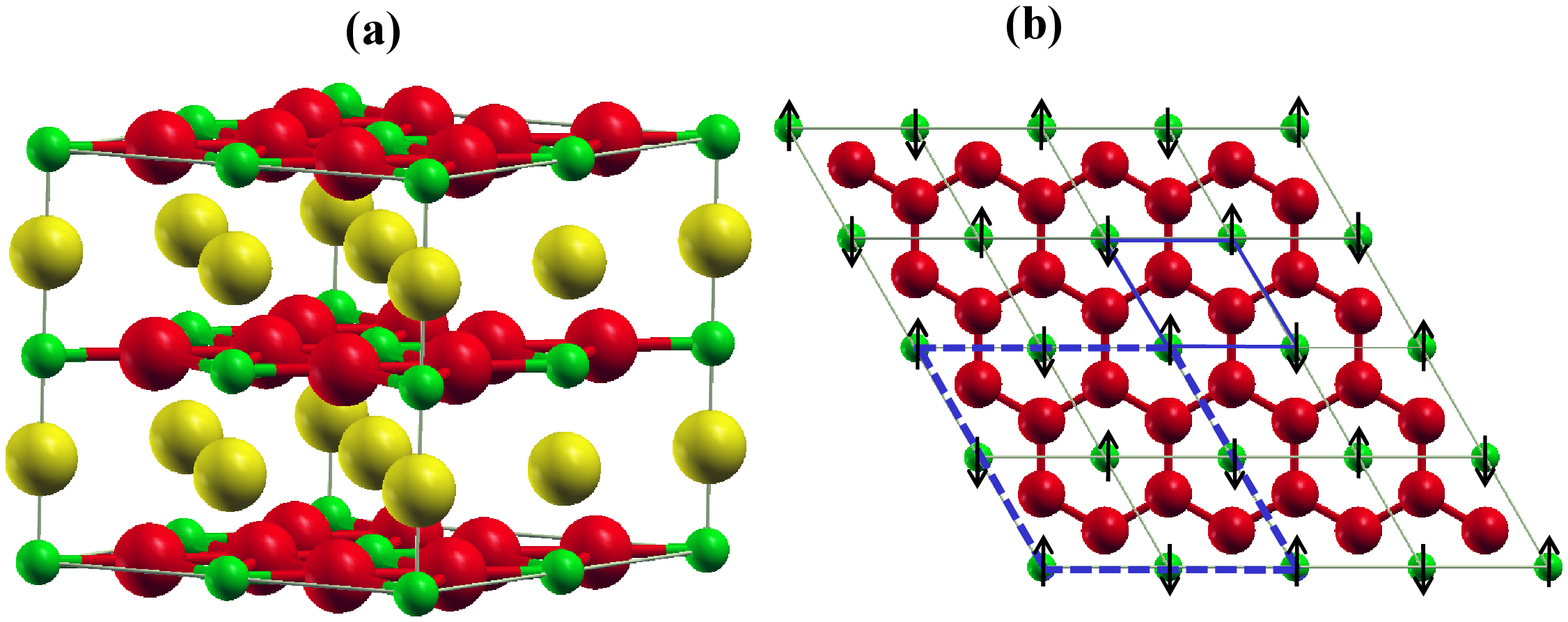}
\caption{(Color online) Crystal structure of bulk Li$_{3}$N ($2\times 2\times 2$) (a) and 2D Li$_{2}$N ($4\times 4\times 1$)
(b). Red and green balls represent Li and N atoms, respectively whereas the yellow balls show
the Li atoms connecting the Li$_{2}$N layers. Solid blue lines represent the unit cell used in the non-
spinpolarized/spin-polarized calculations, whereas the dashed blue lines show the unit cell used
for FM and AFM calculations. Black arrows indicate the direction of the magnetic moments of N
atoms.}
\label{structure}
\end{figure}

Having addressed the atomic origin of magnetism, 
next we investigate 
the stability of magnetic state against changes of the magnetic moments. 
Fig.~\ref{PDOS}(d) shows the scan of 
total energy calculated at different magnetic moments $E(m)$. Clearly, magnetic Li$_{2}$N with
magnetic moment of $1.0\,\mu_{\rm{B}}$ has lower energy (0.16~eV) than the non-magnetic state, and hence
spontaneous magnetism is expected in Li$_{2}$N without any impurities. Note that magnetism
in Li$_{2}$N is confirmed using both LSDA and GGA with the QE code. From the Stoner model we were able 
to predict magnetism in the Li$_{2}$N monolayer and the local magnetic moments carried by
the N atoms. 
To see how these magnetic moments couple with each other and to estimate
the strength of magnetic coupling $J$, the ferromagnetic (FM) and antiferromagnetic (AFM)
coupling between the N atoms in the Li$_{2}$N is investigated by considering a $2\times 2\times 1$ supercell [see Fig.~\ref{structure}(b)]. We found that FM state is more stable than the AFM state by
$\Delta E = E_{\rm{AFM}}-E_{\rm{FM}}=30.5$~meV/unit cell, where $E_{\rm{AFM}}$($E_{\rm{FM}}$) is the total energy per supercell
in the AFM (FM) state. {Note that $\Delta E=108$~meV for buckled Li$_{2}$N.} To estimate the Curie temperature $T_{\rm{c}}$, we map the energetics of
2D Li$_{2}$N onto an Ising model Hamiltonian on a triangular lattice with nearest neighbor interactions
between the magnetic spins [see Fig.~\ref{structure}(b)]. The Ising Hamiltonian $\hat{H}= -\large{\sum_{i,j}JS_{i}S_{j}}$, where
$S_{i,j}$ represents Ising ($\pm 1$) spin on lattice sites and $J$ is the exchange coupling parameter that
gives the nearest neighbor interaction. We estimated $J$ as $J=\Delta E/8 $\cite{delta}, which is {3.81 (13.5)~meV for planar (buckled) Li$_{2}$N.}
The analytic solution of the 2D Ising model on a triangular lattice gives $ k_{\rm{B}}T_{\rm{c}}/J = 3.65$.\cite{47ref}
{Thus our DFT estimated $T_{\rm{c}}$ is about 161 (572)~K for planar (buckled) Li$_{2}$N, and therefore ferromagnetism in Li$_{2}$N monolayer is expected. Hence, buckling in Li$_{2}$N enhances ferromagnetism.} The Li
spin density contours in the FM and AFM states are shown in Fig.~\ref{spin}(b,c) and the spin
density is mostly localized around the N atoms and no spin-polarization at the Li site is
visible. The spin density has $p_{z}$-like character and the N atoms are interacting directly, 
%and the $p_{z}$ electrons can hop through double exchange mechanism\cite{48ref} 
 favouring FM coupling over AFM coupling. The electronic structure of AFM Li$_{2}$N revealed (not shown here)
metallic behavior and this metallicity destabilizes the AFM phase, and half-metallic FM
Li$_{2}$N becomes the ground state of Li$_{2}$N.

{Before we summarize our DFT calculated results, it will be necessary to consider the dynamical stability, which can be addressed through phonon dispersion curve.  We calculated the phonon dispersion curve of a planar 2D-Li$_{2}$N (not shown here) and we found imaginary frequencies at the $\Gamma$, $K$ and $M$-points in the BZ, and  such imaginary frequencies indicate the lattice instability in Li$_{2}$N. Note that similar imaginary frequency is also observed in a planar silicene and germanene.\cite{sili2d} Interestingly the lattice instability is removed in the buckled Li$_{2}$N. Fig.~\ref{PDOS}(e) shows the calculated phonon dispersion curve of buckled Li$_{2}$N, and it is clearly showing that the buckled Li$_{2}$N is dynamically stable, and no imaginary frequency can be seen. 
Therefore, we can conclude that buckling not only stabilizes Li$_{2}$N but also increases the $T_{\rm{c}}$. Further experimental work would be needed to confirm our computational predictions.}

\section{Summary}
To summarize, we used DFT calculations to propose a new 2D magnetic material Li$_{2}$N. The electronic structure of bulk Li$_{3}$N is also investigated and we observed that bulk Li$_{3}$N is a nonmagnetic semiconductor. However, the monolayer Li$_{2}$N showed interesting signature in the electronic density of states. The density of states at the Fermi energy in 2D Li$_{2}$N was increased in the non-magnetic calculations. The large density of states was discussed in terms of low coordination number. The narrow band at the Fermi energy gave an indication of possible magnetism in Stoner's mean field model. The spinpolarized (magnetic) calculations showed that 2D Li$_{2}$N is magnetic without any defects -- this fact is confirmed by using both the SIESTA and QE codes using LSDA and GGA. The electronic band structure of 2D Li$_{2}$N was observed to be a half-metal. The DFT calculated magnetic moment of Li$_{2}$N was found to be 1.0 $\mu_{\rm{B}}$ per unit cell. The projected density of states showed that the local magnetic moments in Li$_{2}$N are contributed by the N-$p$ orbitals. 
{In contrast to planar Li$_2$N, buckled Li$_2$N is also dynamically stable, as evidence by its phonon modes.} 
{Finally, to estimate the Curie temperature $T_{\rm{c}}$ of our new proposed material Li$_{2}$N, we also considered the FM and AFM coupling between the N atoms both in the planar and buckled Li$_{2}$N. The DFT computed data was used in the Ising model, and we showed that the $T_{\rm{c}}$ of planar (buckled) Li$_{2}$N is about 161 (572)~K.}

\newpage

\newpage

\newpage

\newpage

\newpage
\begin{thebibliography}{99}
  
\bibitem{Sarma} S. D. Sarma, S. Adam, E. H. Hwang, and E. Rossi,
Rev. Mod. Phys. 83, \textbf{407} (2011).


\bibitem{Xu}M. S. Xu, T. Liang, M. M. Shi, and H. Z. Chen,
Chem. Rev.113, 3766 (2013).

\bibitem{Mak}K. F. Mak, C. Lee, J. Hone, J. Shan, and T. F. Heinz, Phys. Rev. Lett. \textbf{105}, 136805 (2010).



\bibitem{Liu}H. Liu, A. T. Neal, Z. Zhu, X. Xu, D. Tomanek, and P. D. Ye, ACS Nano \textbf{8}, 4033 (2014).


\bibitem{Cai}B. Cai, S. L. Zhang, Z. Y. Hu, Y. H. Hu, Y. S. Zou, and H. B. Zeng, Phys. Chem. Chem. Phys. \textbf{17}, 12634 (2015).




\bibitem{elec} E. Canadell, Chem. Mater. \textbf{10}, 2770 (1998).

\bibitem{book}K. Barnham and D. D. Vvedensky, \textit{Low-Dimensional Semiconductor Structures: Fundamentals
and Device Applications}, Cambridge University Press (2001).

\bibitem{16ref} A. Neto, F. Guinea, N. Peres, K. Novoselov, A. Geim, Rev. Mod. Phys. \textbf{81}, 109 (2009).

\bibitem{17ref} G. Rahman, Euro, Phys. Lett. \textbf{105},  37012 (2014).
\bibitem{18ref} S. Lebegue and O. Eriksson, Phys. Rev. B \textbf{79}, 115409 (2009).
\bibitem{19ref} K. Takeda and K. Shiraishi, Phys. Rev. B \textbf{50}, 14916 (1994).
\bibitem{20ref} N. D. Drummond, V. Zolyomi, and V. I. Falk, Phys. Rev. B \textbf{85}, 075423 (2012).
\bibitem{21ref} E. Durgun, S. Tongay, and S. Ciraci, Phys. Rev. B \textbf{72}, 075420 (2005).
\bibitem{22ref} M. Zhang et al., Chem. Phys. Lett. \textbf{379}, 81 (2003).
\bibitem{23ref} S. Cahangirov, M. Topsakal, E. Akturk, H. Sahin, and S. Ciraci, Phys. Rev. Lett. \textbf{102}, 236804 (2009).

\bibitem{Kane}N. D. Drummond, V. Z\'{o}lyomi, and V. I. Fa\'{l}ko, Phys. Rev. B \textbf{85}, 075423 (2012).



\bibitem{Cr2017}A. V. Kuklin, A. A. Kuzubov, E. A. Kovaleva, N. S. Mikhaleva, F. N. Tomilin, H. Lee and P. V. Avramov,  Nanoscale, \textbf{9}, 621 (2017).

\bibitem{Cr2017Li}J. Liu, Z. Liu, T. Song and X. Cui,  J. Mater. Chem. C, \textbf{5}, 727 (2017).

\bibitem{Cr2016}J. Liu, Q. Sun, Y. Kawazoe and P. Jena, Phys Chem Chem Phys. \textbf{18}, 8777 (2016).

\bibitem{Cr2017H}B. Huang, G. Clark, E. Navarro-Moratalla, D. R. Klein, R. Cheng, K. L. Seyler, D. Zhong, E. Schmidgall, M. A. McGuire, D. H. Cobden, W. Yao, D. Xiao, P. Jarillo-Herrero and X. Xu, \textit{et~al.} Nature \textbf{546}, 270 (2017)



\bibitem{28ref}S. Blundell, \textit{Magnetism in Condensed Matter} (Oxford Master Series in Condensed Matter
Physics) (2001).


\bibitem{mot} M. Cyrot, J. Phys. (Paris) \textbf{33}, 125 (1972).
\bibitem{26ref} A. Ur Rahman, G. Rahman, and 
{V. M. Garc\'{\i}a-Su\'arez}, J. Magn. Magn. Mater. \textbf{443}, 343 (2017).





\bibitem{ref1} A. Rabenau, in Festk{\"o}rperprobleme (Advances in Solid State Physics) edited by J. Treusch (Vieweg, Braunschweig, 1978), vol. 18, pp. 77-108.
\bibitem{ref2} P. Chen, Z. Ziong, J. Luo, J. Lin, and K. Lee Tan, Nature (London) \textbf{420}, 302 (2002). 
\bibitem{ref3}T. Ichikawa, S. Isobe, N. Hanada, and H. Fujii, J. Alloys Compd. \textbf{365}, 271 (2004). 

\bibitem{ref5}Y. Nakamori, G. Kitahara, K. Miwa, S. Towata, and S. Orimo, Appl. Phys. A: Mater. Sci. Process. \textbf{80}, 1 (2005).

\bibitem{35ref}P. Novak and F. R. Wagner, Phys. Rev. B \textbf{66}, 184434 (2002).

\bibitem{37ref} V. P. Antropov and V. N. Antonov, Phys. Rev. B \textbf{90}, 094406 (2014).
\bibitem{38ref} A. Jesche, L. Ke, J. L. Jacobs, B. Harmon, R. S. Houk, and P. C. Canfield, Phys. Rev. B \textbf{91}, 180403(R) (2015).






\bibitem{ref8} H.J. Beister, S. Haag \textit{et al}; Angew. Chem, Int. Ed.
Engl. \textbf{27}, 1101 (1988).




\bibitem{DFT} P. Hohenberg and W. Kohn, Phys. Rev. \textbf{136}, B846 (1964).

\bibitem{siesta} J. M. Soler et al J. Phys.Condens.: Matter \textbf{14}, 2745 (2002).

\bibitem{CM05} P. Giannozzi, S. Baroni, N. Bonini, M. Calandra, R. Car, C. Cavazzoni, D. Ceresoli, D. L. Chiarotti, M. Cococcioni, I. Dabo, A. D. Corso, S. de Dironcoli, S. Fabris, G. Fratesi, R. Gebauer, U. Gerstmann, C. Gougoussis, A. Kokalj, M. Lazzeri, L. Martin-Samos, N. Marzari, F. Mauri, R. Mazzarello, S. Paolini, A. Pasquarello, L. Paulatto, C. Sbraccia, S. Scandolo, G. Sclauzero, A. P. Seitsonen, A. Smogunov, P. Umari, R. M. Wentzcovitch, J. Phys. Condens. Matter {\bf 21}, 39 (2009).




\bibitem{lda} J. P. Perdew  and A. Zunger, Phys. Rev. B \textbf{23}, 5048 (1981).



\bibitem{CM02} D. V. Rybkovskiy, N. R. Arutyunyan, A. S. Orekhov, I. A. Gromchenko, I. V. Vorobiev, A. V. Osadchy, E. Y. Salaev, T. K. Baykara, K. R. Allakhverdiev, and E. D. Obraztsova, Phys. Rev. B \textbf{84}, 085314 (2011).



\bibitem{CM04} J. P. Perdew, K. Burke and M. Ernzerhof, Phys. Rev. Lett. \textbf{77}, 3865 (1996).



F. D. Murnaghan, Proc. Natl Acad. Sci. USA \textbf{30}, 244 (1944).

\bibitem{lctheory} Y. Yan, J.Y. Zhang, T. Cui, Y. Li, Y.M. Ma, J. Gong, Z.G. Zong, and G.T. Zou, Eur. Phys. J. B \textbf{61}, 397 (2008).

\bibitem{45ref}H. Brendecke and W. Bludau, J. Appl. Phys. \textbf{50}, 4743 (1979).



\bibitem{39ref} S. Wu, Z. Dong, F. Boey, and P. Wu, Appl. Phys. Lett. \textbf{94}, 172104 (2009).
\bibitem{40ref} G. Rahman and H. U Jan, J. Supercond. Nov. Magn. (2017), doi 10.1007/s10948-017-4224-0

%\bibitem{41ref} G. Rahman and S. Sarwar, J. Supercond. Nov. Magn. (2017),doi 10.1007/s10948-017-4223-1

\bibitem{bccfe} G. Liu, D. N. Manh, B. G. Liu, and D. G. Pettifor, Phys. Rev. B \textbf{71}, 174115 (2005).

\bibitem{Topsakal}M. Topsakal, S. Cahangirov, E. Bekaroglu, and S. Ciraci, Phys. Rev. B \textbf{80}, 235119 (2009).

 
%PK if it's not relevant, we don't cite it
%\bibitem{zhang2009}C.-w. Zhang, S.-s. Yan, Appl. Phys. Lett., 95, 232108 (2009).

\bibitem{zhang2012}C.-w. Zhang, S.-s. Yan, J. Phys. Chem. C, 116, 4163 (2012).

\bibitem{zhang2012NRL}F.-b. Zheng, C.-wen Zhang,Nanoscale Research Letters, 7, 422 (2012).
%PK if it's not relevant, we don't cite it
%\bibitem{wang2017}Wang et.al, Appl. Phys. Lett., 110, 233107 (2017).

\bibitem{42ref} W. Zhong, G. Overney, and D. Tomanek, Phys. Rev. B \textbf{47}, 95 (1993).


\bibitem{yambo}A. Marini, C. Hogan, M. Gruning, D. Varsano,
Comp. Phys. Commun. \textbf{180}, 1392 (2009).

\bibitem{delta}$E_{\rm FM}= -6J, \, E_{\rm AFM}=4J-2J, \, \Rightarrow\,\Delta\,E=8J$, since the spin of 4 (out of 6) neighbors has been turned in the AFM structure. 

\bibitem{47ref} M. E. Fisher and R. J. Burford, Phys. Rev. \textbf{156}, 583 (1967).


\bibitem{sili2d} S. Cahangirov, M. Topsakal, E. Akturk, H. {\,S}ahin, and S. Ciraci, Phys. Rev. Lett. \textbf{102}, 236804 (2009).
%\bibitem{delta}$FM=4J_{1}+2J_{2}, AFM=-4J_{1}+2J_{2},\Rightarrow\,\Delta\,E=8J_{1}$, where $J_{1}$ and $J_{2}$ are the first and second nearest neighbor exchange constant, respectively. We used $J_{1}=J$.
%PK changed 
 
\end{thebibliography}
\end{document}